# Boosted top quark tagging and polarization measurement using machine learning

Soham Bhattacharya[*]

*Deutsches Elektronen-Synchrotron (DESY), Notkestraße 85, Hamburg 22607, Germany*

Monoranjan Guchait[†] and Aravind H. Vijay[‡]

*Department of High Energy Physics, Tata Institute of Fundamental Research (TIFR), Homi Bhabha Road, Mumbai 400005, India*

Machine learning techniques are used for treating jets as images to explore the performance of boosted top quark tagging. Tagging performances are studied in both hadronic and leptonic channels of top quark decay, employing a convolutional neural network (CNN) based technique along with boosted decision trees (BDT). This computer vision approach is also applied to distinguish between left and right polarized top quarks. In this context, an experimentally measurable asymmetry variable is proposed to estimate the polarization. Results indicate that the CNN based classifier is more sensitive to top quark polarization than the standard kinematic variables. It is observed that the overall tagging performance in the leptonic channel is better than the hadronic case, and the former also serves as a better probe for studying polarization.

## I. INTRODUCTION

In the Standard Model (SM) of particle physics the top quark plays an important role by its own virtue. For instance, how the large mass of the top quark is offered by the SM Higgs sector, is one of the most important and fundamental questions in electroweak symmetry breaking. Moreover, in several beyond the Standard Model (BSM) signal processes, it is envisaged that heavy non-SM particles produced at the TeV energy scale can access the phase space where they can decay to a final state consisting of very energetic (boosted) top quarks. The decay products from such boosted top quarks are highly collimated, and thus form a single "fat" jet (hereon referred to as a fatjet), instead of multiple jets. Hence, there is a great demand for developing strategies to tag boosted objects as a single fatjet, preserving all its original properties. Several techniques are already available for tagging boosted top jets, which are widely used in searches of new physics models having very massive [$\sim\mathcal{O}(1 \text{ TeV})$] particles [1–7]. While tagging the top jet in its hadronic decay channel is found to be very popular and well studied, interest in tagging the leptonic channel is also growing [8]. Needless to say, more dedicated studies are required to tag boosted leptonic top jets with better techniques.

Furthermore, another important aspect of top quark physics is the measurement of its polarization, which can potentially serve as an alternate avenue to probe the existence and nature of BSM physics coupled with the top quark sector. It is well known that the kinematic properties (in particular, angular distributions) of the top quark decay products are guided by its polarization, which is essentially decided by the structure of the coupling at the top quark production vertex [9]. In previous studies, the lepton has been found to be a suitable object to study the polarization of less energetic top quarks [10]. However, such methods are not as effective in the case of boosted top quarks because of the overlapping of its decay products in the kinematic phase space [11–14]. Recently, strategies for measuring the polarization of boosted top quarks in the hadronic decay channel have been proposed in Ref. [14] for the LHC experiments.

Presently a great deal of interest has been evolving in the application of sophisticated machine learning (ML) techniques in analyzing physics events, as well as object reconstruction and identifications in high energy physics experiments (more details can be found in this review [5]). The use of ML techniques has proven to be very powerful in improving the significance through better classification of signal and background categories, where cut based analyses are found to be comparatively less effective.

[*]soham.bhattacharya@cern.ch
[†]guchait@tifr.res.in
[‡]aravindhv10@gmail.com

The key feature of one such popular ML technique, is to represent a jet as an image [15–19]. A jet image is essentially a presentation of the energies or transverse momenta ($p_T$) of its constituents in a certain plane divided into $N \times N$ cells, called pixels. Generally, low level object information (e.g., track momenta, calorimeter cell energies, etc.) is used instead of high level objects to form the jet images. The energy or momentum deposited in each pixel, defined as the pixel strength, is used as the input to train an neural network (NN) which learns to distinguish between various classes of objects [19,20]. Among several existing NNs, the CNN is widely employed for classifying jets at the LHC and also has been used in this study.

The primary goal of this analysis is two fold. In the first step, the tagging performance of boosted top quarks is studied for both its hadronic and leptonic decay channels using jet images as input to the CNN. Further, the classification performances are improved by using the CNN output as one of the inputs to a boosted decision tree (BDT). Revisiting the hadronic top tagging performances for completeness, we mainly focus on the leptonic decay mode, which we believe to the best of our knowledge, has not been studied in detail using ML techniques. The advantage of this image based leptonic top tagging technique is that it does not require the lepton to be well identified, which is a nontrivial task due to the presence of high hadronic activity around it inside a boosted top fatjet. In this leptonic top tagging study, we consider a wide variety of boosted jets as background categories, namely, light flavor QCD, hadronic top, and $W/Z$ bosons (both hadronic and leptonic decay modes). In general, it is found that the overall tagging performance is comparatively better for the leptonic than the hadronic channel of top quark decay. In the second half of the paper, we employ the CNN based technique to differentiate between left and right polarized boosted top quarks, in both hadronic and leptonic decay channels. In order to quantify the polarization of boosted top quarks, an experimentally measurable asymmetry variable is constructed, which resembles the asymmetry observable proposed in Ref. [14]. This asymmetry variable is then used to study and compare the performance of the CNN based technique to a few other kinematic observables for polarization [14].

It is to be noted that recent studies have demonstrated that graph network based taggers such as ParticleNet [21] and LundNet [22] perform somewhat better than CNN based techniques. The performance of these techniques in the context of leptonic top tagging and polarization can be an interesting study—however this is beyond the scope of this paper, and we postpone it for a future study.

The paper is organized as follows. Describing the methodology in Sec. II, the top tagging performances are discussed in Sec. III. The effects of top polarization on jet images, and the corresponding performances of the CNN in distinguishing left and right handed top quarks are discussed in Sec. IV. Finally, the results are summarized in Sec. V.

## II. METHODOLOGY

Boosted top quarks are produced by generating top pair ($pp \to t\bar{t}$) and $W'$ ($pp \to W' \to tb$) events, where the $W'$ mass is set to 3 TeV. Here on we refer them as $t\bar{t}$ and $W'$ event samples, respectively. Light flavor jets produced in hard QCD events are treated as a background for boosted top jets. Boosted $W/Z$ bosons produced in $W/Z +$ jets events are also included as a source of background.

For our polarization study, left and right polarized top quarks are generated via the process, $W' \to tb$, where the interaction Lagrangian reads as [23],

$$\mathcal{L}_{W'tb} \sim f_i \gamma^\mu [g_R(1+\gamma_5) + g_L(1-\gamma_5)] W'_\mu f_j. \quad (1)$$

The left (right) polarized top quarks from $W'$ decay can be produced by explicitly setting the coupling strength $g_R = 0$ ($g_L = 0$). Another set of polarized top quarks are produced in a supersymmetric process, where lighter top squark pair events ($pp \to \tilde{t}_1 \bar{\tilde{t}}_1$) are generated. The top squark mass is set to 1 TeV and it is forced to decay to a top quark and a lightest neutralino ($\tilde{\chi}_1^0$) of mass 100 GeV. Again, the chirality of this top quark can be controlled by appropriately adjusting the couplings in the interaction Lagrangian [14],

$$\mathcal{L}_{t\tilde{t}_1 \tilde{\chi}_1^0} = \bar{\tilde{\chi}}_1^0 (g_L^{\tilde{t}_1} P_L + g_R^{\tilde{t}_1} P_R) \tilde{t}_1 + \text{H.c.} \quad (2)$$

Hence, by suitably choosing the neutralino composition to be pure (100%) gaugino or Higgsino like and the top squark sector mixing angle, one can make $g_L^{\tilde{t}_1} \gg g_R^{\tilde{t}_1}$ and vice versa. We refer to this as the SUSY sample henceforth. The $t\bar{t}$ and QCD samples are generated using PYTHIA 8 [24], and the $W/Z +$ jets sample using MadGraph5_aMC@NLO [25]. In order to access the boosted region of the phase space, a cut of 400 GeV is applied to the $p_T$ of the outgoing partons at tree-level ($\hat{p}_{T,\min} = 400$ GeV) for the $t\bar{t}$, QCD, and $W/Z +$ jets processes. The $W'$ sample is produced interfacing FeynRules v2.0 [26] in the framework of an effective theory with MadGraph5_aMC@NLO. The SUSY sample is also generated using MadGraph5_aMC@NLO. Eventually, all the aforementioned processes are hadronized using PYTHIA 8, with full multiparton interactions turned on. The detector simulation for each sample is performed by passing the generated events through DELPHES v3.4 [27] with its compact muon solenoid (CMS) card.

It is to be noted that the effect of pileup is not taken into account in the analysis, as detailed pileup simulation is out of the scope of this study due to limited computing resources. Intuitively, the impact of pileup on our results involving very high momentum objects is not expected to be severe, as the effect of pileup is typically greater for low or moderately energetic objects. Additionally, sophisticated techniques are available to mitigate the effect of pileup, producing results that are fairly robust against pileup [28].

Using the DELPHES EFlow objects (namely, EFlowTrack, EFlowPhoton, and EFlowNeutralHadron), fatjets of radii $R = 1.5$ are reconstructed with the anti-$k_T$ [29,30] jet algorithm. The fatjets are selected with a cut of $p_T > 200$ GeV and pseudorapidity $|\eta| < 2.4$. A fatjet is categorized as a hadronic (leptonic) top jet if the jet axis lies within a cone of $\Delta R = \sqrt{\Delta y^2 + \Delta \phi^2} < 1.0$ around the resultant momentum of the generator-level visible decay products of a hadronically (leptonically) decaying top quark. Similarly, in $W/Z$ + jets events, a leptonic (hadronic) $W/Z$ jet refers to a fatjet which has been matched, with the aforementioned $\Delta R$ criterion, to the resultant momentum of the generator-level visible decay products of a leptonically (hadronically) decaying $W/Z$ boson. Here $\Delta y$ and $\Delta \phi$ represent the differences in rapidities and azimuthal angles of the two objects in consideration. The fatjets have been cleaned using the soft-drop procedure [31] with $\beta = 0$ and $z_{\text{cut}} = 0.1$, which is one of the standard configurations used at the CMS experiment [28]. The aforementioned jet clustering and cleaning have performed using FastJet v3.2.1 [30].

Jet images are preprocessed with certain transformations to aid the network in learning their features, so as to improve the classification performance. The preprocessing procedure follows the methodology described in Ref. [32]. In this technique, the momentum of a jet is rescaled such that its new mass has a fixed value $m_B$, and boosted to a frame where its energy has a constant value $E_B$. Consequently, the boost factor $\gamma_B = E_B/m_B$ assumes a fixed value irrespective of the initial momentum of the jet. The added advantage of this preprocessing technique is that the transformed jets are neither too heavily boosted nor boosted at all, since either scenario is not expected to be well suited for image based classification. It is to be noted that the ratio $\gamma_B$ is the physical parameter of import, and not the individual values of $m_B$ and $E_B$. In this study, we set $\gamma_B = 2$ [32], ensuring the subjets are well resolved without any loss of generality. This makes our training less sensitive to the original boost of the top quark, and thus applicable to a wide range of top quark energies. A Gram-Schmidt transformation is applied to the jet such that the image plane is perpendicular to the jet axis and the two subjets with the highest energies lie along the $x$-axis of the image plane. Note that the jet is required to have at least three constituents for this procedure to work. The image of a jet is a $50 \times 50$ histogram whose cells/pixels are filled with the fraction of the jet's energy carried by each its of constituents.

Figure 1 shows the images of preprocessed hadronic and leptonic top jets from $t\bar{t}$ (top panel), hadronic and leptonic $W$ jets (middle panel), leptonic $Z$ jets, and light flavor QCD jets (bottom panel). In general, a bright spot mainly from the leading subjet is observed, along with a less sharp spot surrounded by a diffused cloud from the subleading subjets. For hadronic top jets, the spot on the left is broader compared to QCD jets. This can be attributed to the fact that the three subjets inside the top jet correspond to three hard partons, whereas for QCD the subjets are arising primarily from soft radiation. For hadronic $W$ jets, two distinct spots are seen, which are due to the two primary jets from $W$ decay. Notice that in this case the two spots are not as connected, which is obviously because of the absence of outward color flow in $W$ decay. For leptonic top jets, two relatively well separated spots are observed, where one is arising from the $b$ quark, and the other from the lepton. Since the b quark and lepton do not have any color connection, no diffused cloud is seen between the leading and subleading spots, unlike the images of hadronic top and QCD jets. In case of leptonic $W$ jets, the bright narrow spot on the right can be attributed to the lepton, where as the diffused spot on the left is due to contamination from hadronic activities (like additional jets) in $W$ + jets events. On the other hand, the image of leptonic $Z$ jets shows two sharp spots which are because of the two leptons in the $Z$ decay. These two spots are significantly narrower unlike the hadronic jets. The CNN is able to distinguish between different categories of jet images by learning these distinct features. We separate each jet into its track, photon, and neutral hadron components, and thus three images for each jet (i.e., three input channels/layers) are used to train the network. Recall that the objects used to construct the jets and their images are the DELPHES EFlow objects. The EFlowTracks are reconstructed using tracker information, and its corresponding layer (image) can be, in principle, binned more finely compared to the calorimeter based layers (i.e., the neutral components). However, the track resolution is more detector specific, and we have used the same granularity for all the layers to keep the applicability of the results more general.

The network architecture used in this study is described in Fig. 2. The hyperparameters (the number of channels, kernel sizes, number of nodes, etc.) have been slightly adjusted in order to obtain a reasonably good compromise between the performance and training time. For the purpose of training the network, we have used the Xavier initialization [33] for the weights and the Adam gradient descent [34] with a batch size of 100 and a learning rate (step size of the gradient descent) of 0.001. We have implemented the aforementioned architecture using the gluon API of Apache MXNet v1.5.1 [35] in python.

### III. TOP TAGGING

Network trainings are performed using about $\sim$1M/1M of signal/background jet images corresponding to the following combinations: (i) hadronic top and QCD jets, (ii) leptonic top and QCD jets, (iii) leptonic and hadronic top jets, (iv) leptonic top and hadronic $W$ jets, (v) leptonic top and hadronic $Z$ jets, (vi) leptonic top and leptonic $W$ jets, and (vii) leptonic top and leptonic $Z$ jets. Approximately $\sim 100K/100K$ of signal/background jet images are used for the purpose of testing to ensure that the network is not overtrained. We have used top jets from the $t\bar{t}$ sample for training. The network is trained for 25 epochs, at which stage the training and testing losses are found to saturate to almost identical values.

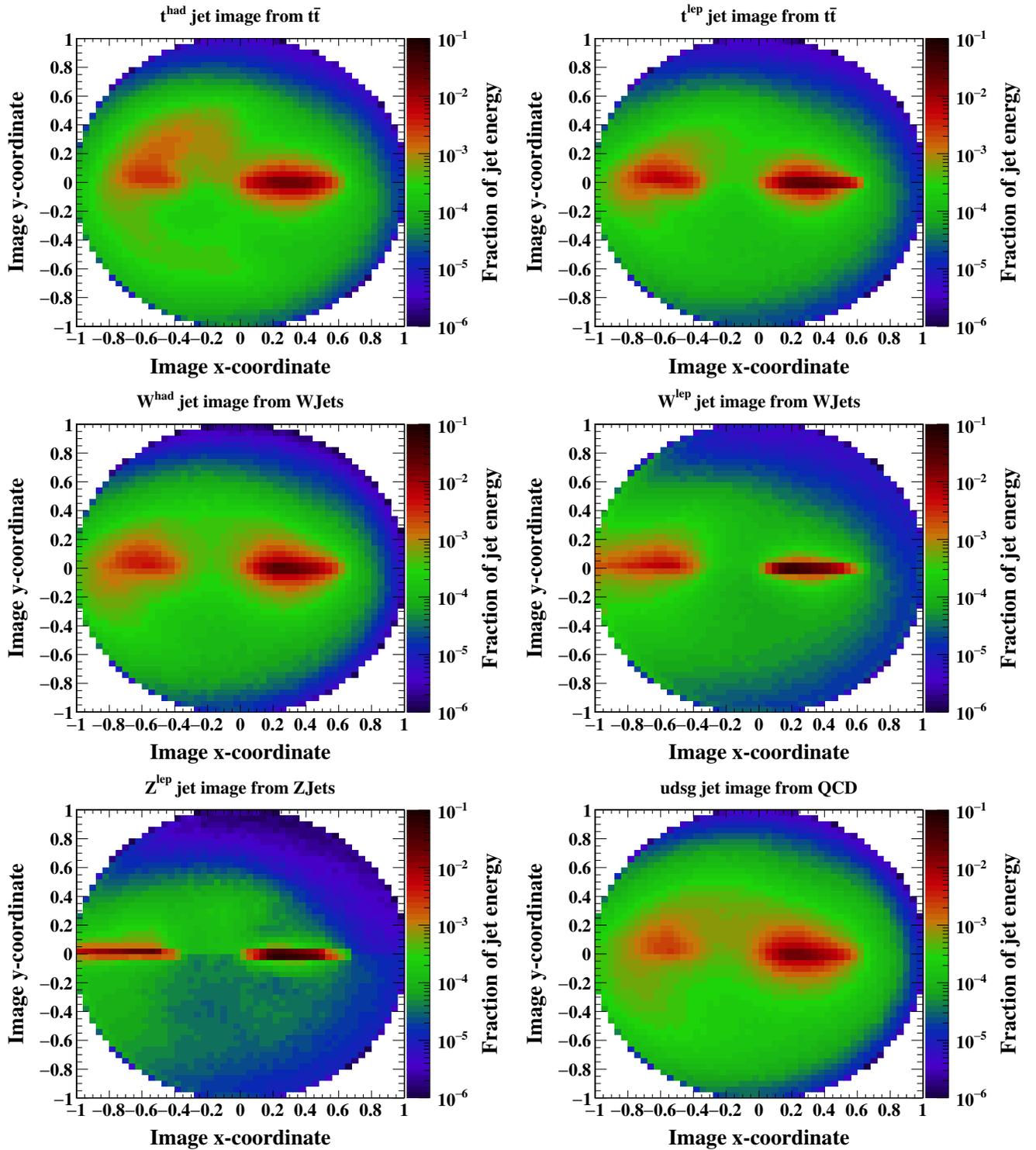

FIG. 1. Images of hadronic (upper left) and leptonic (upper right) top jets from $t\bar{t}$ events, hadronic (middle left) and leptonic (middle right) $W$ jets from $W+$ jets events, leptonic $Z$ jets (lower left) from $Z+$ jets events, and light flavor QCD jets (lower right). The image of hadronic $Z$ jets looks very similar to that of hadronic $W$, and is hence not presented here. These are the inclusive images of jets where the track, photon and neutral hadron components have been combined.

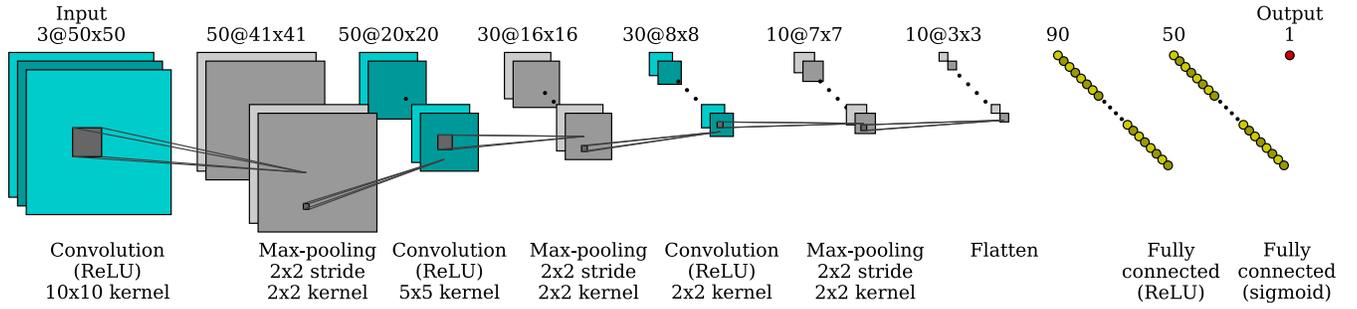

FIG. 2. A schematic diagram of the network structure. For any given layer, the text above it indicates the shape of the layer. The shape of a convolution/max-pooling layer (in cyan/gray squares) is represented as channels@ $N \times N$. For a fully-connected layer (in yellow circles) it is a single number corresponding to its number of nodes. The text at the bottom indicates the details of the operation performed on the layer above it in order to obtain the next layer. This includes the kernel sizes used for the convolution and the max-pooling operations, along with the activation function (ReLU/sigmoid). This diagram has been generated by adapting the code from [36].

Figure 3 shows the receiver operating characteristic (ROC) curves to illustrate the hadronic/leptonic top jet discrimination against QCD jets, in solid red/blue (left panel). Clearly, leptonic top jets are better tagged than the hadronic ones. This can be attributed to the comparatively clean environment of the leptonic decay mode, which is evident from the jet images presented in Fig. 1. The robustness of the CNN trainings is also tested on top jets from the aforementioned $W'$ sample, and the corresponding performances are presented in dashed lines in Fig. 3 (left panel). The differences between the ROC curves for $t\bar{t}$ and $W'$ samples are primarily due to higher $p_T$ of top jets originating from massive $W'$. It is important to note that we have checked that without the jet preprocessing discussed in the previous section, the observed differences are far more significant. So although not perfect, the preprocessing greatly helps to reduce the dependence of the result on the initial boost of the top quark. The leptonic top tagging performance is also studied by training the CNN with hadronic top jets from $t\bar{t}$ as background, and the corresponding performances in the $t\bar{t}$ and $W'$ samples are presented in Fig. 3 (right). The difference in performances between the two samples is observed to be small.

The performance of top tagging in its leptonic channel is further investigated by training the network with boosted leptonic and hadronic $W/Z$ jets (from the $W/Z$ + jets samples) as background. The ROC curves are presented in Fig. 4 where the performance of leptonic top tagging is shown against leptonic (left panel) and hadronic (right panel) $W/Z$ jets as background. The performance is also tested using leptonic top jets originating from the $W'$ sample, and are shown in the same figure. As we see in Fig. 1, the presence of two leptons in leptonic $Z$ jets makes the corresponding jet image very clearly distinguishable from that of leptonically decaying top quarks. Consequently, the training against the leptonic $Z$ jet background is comparatively better than that against leptonic $W$. However, it is to be noted that only nonisolated leptons from leptonic $W$ jets are treated as background here (as the fatjet is required to have at least three constituents). The lepton in most boosted $W$ decays will be highly isolated, and can be rejected in the first place with very simple isolation criteria. The tagging performance with hadronic $W$ jets as the background (right panel) indicates that leptonic top jets from the $W'$ sample are slightly better tagged than those from the $t\bar{t}$ sample. Note that the performance against hadronic $Z$ jets is not shown as it is very similar to that against hadronic $W$ jets.

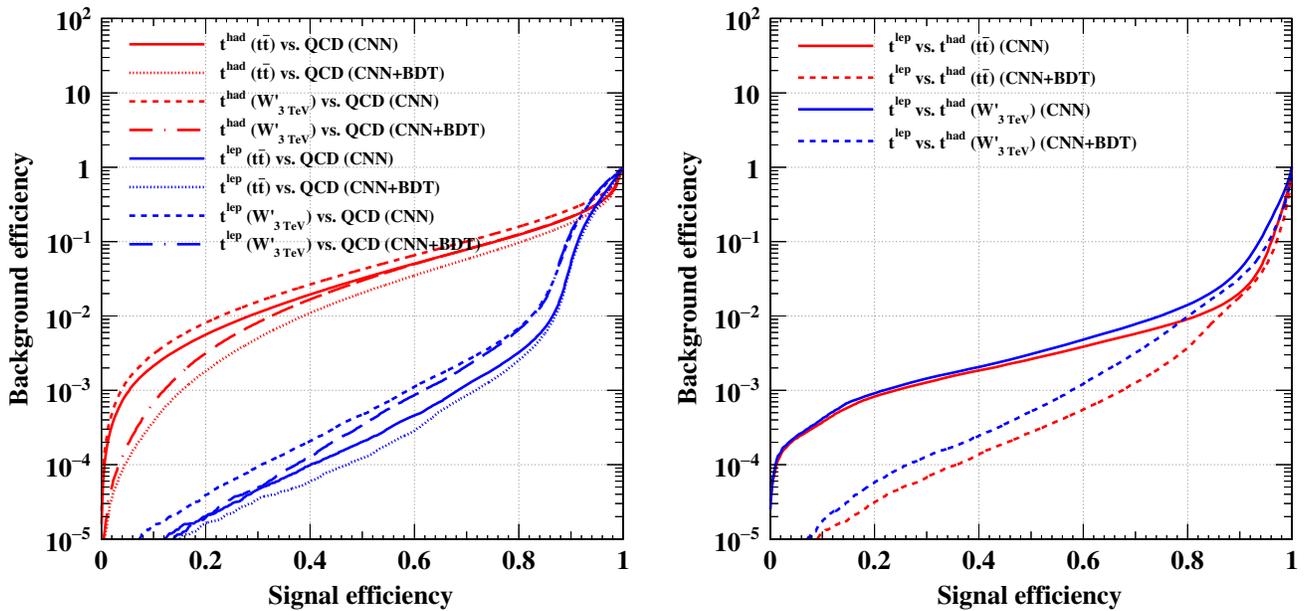

FIG. 3. The left figure shows the ROC curves corresponding to the hadronic (leptonic) top versus QCD jet trainings in red (blue). The right figure shows the leptonic top tagging ROC using hadronic top jets as the background. The performance of the trainings when evaluated on top jets from the $W'$ sample, are also presented in both the figures.

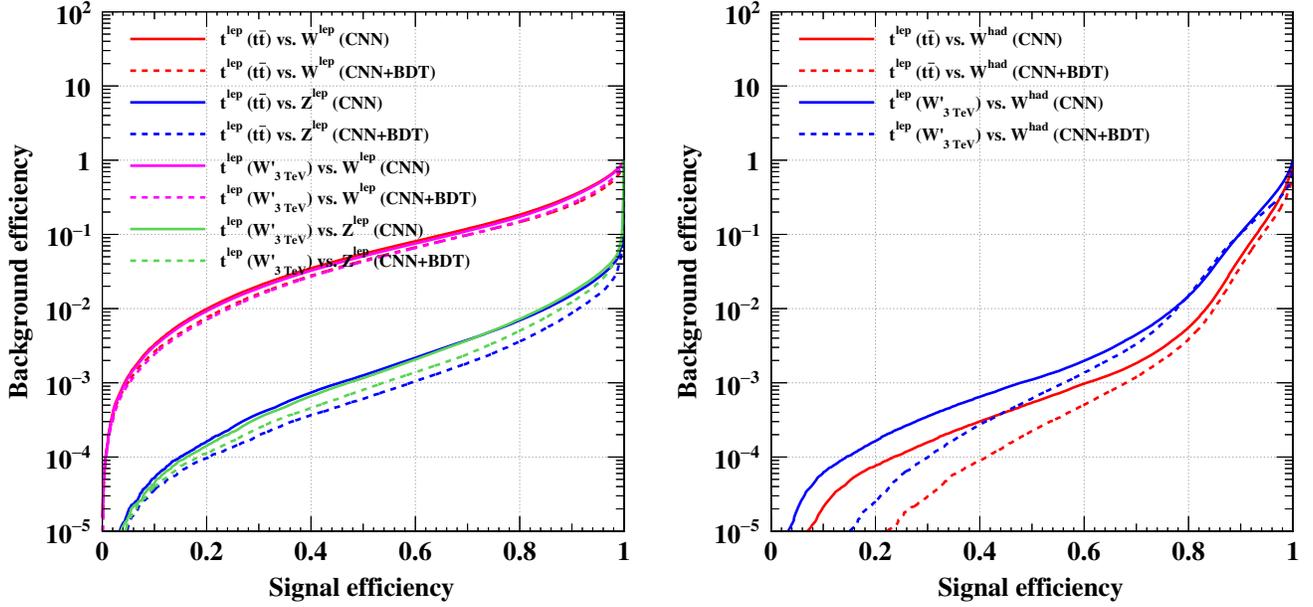

FIG. 4. The left figure shows the ROC curves corresponding to the trainings for leptonic top versus leptonic $W$ and $Z$ jets. The right figure shows the leptonic top tagging ROC using hadronic $W$ jets as the background. The network trained using hadronic $Z$ jets as the background performs very similarly, and is thus not presented here. The performance of the trainings when evaluated on top jets from the $W'$ sample, are also presented in both the figures.

We try to further improve the obtained tagging performances for both hadronic and leptonic channels by training a BDT implemented in the TMVA [37] framework, where the training and testing samples are the same as that used for the CNN. In this training, the additional variables (apart from CNN classifier) used are the jet mass ($m_j$) and the ratios of the N-subjettiness variables such as, $\tau_2/\tau_1$, $\tau_3/\tau_2$, and $\tau_4/\tau_3$ [38]. The $\tau_N$ variable is defined as,

$$\tau_N = \frac{1}{R_0 \sum_k p_{T,k}} \sum_k p_{T,k} \min(\Delta R_{1,k}, R_{2,k}, \ldots R_{N,k}). \quad (3)$$

Here $\Delta R_{i,k}$ are the geometrical separations between $i$th subjet and $k$th subjettiness axes, and $R_0$ is the jet radius. The BDT based performances (labeled CNN + BDT) are presented along with the CNN performances in both the Figs. 3 and 4. Even in this case the robustness of the training is verified by evaluating the BDT trained using top jets from $t\bar{t}$ events, on those from the $W'$ sample. An overall improvement over the CNN training is observed in almost for all cases. For instance, the CNN + BDT discrimination against QCD jets is better by a factor of $\approx 1.5$–2 at 40% signal efficiency. On the other hand, the leptonic versus hadronic top jet discrimination improves almost by a factor of $\approx 10$ at the same signal efficiency.

Lastly, it is possible to add lepton specific information in the BDT to further improve the tagging performance against hadronic jet backgrounds. However, the construction of such variables (such as shower-shape variables in the calorimeter for electrons, and information from the muon spectrometer) is highly detector specific, and thus more appropriately done with a realistic detector simulation.

## IV. TOP POLARIZATION

In this section we study the measurement of boosted top quark polarization using jet images, and then compare the performance with the typical kinematic polarimeter variables [10,14]. Recall that the angular distributions of the top quark decay products are governed by the structure of its coupling [39]. It is worth noting here that in case of hadronic decaying top quark, the d-type quark from $W$ decay carries the maximum spin analyzing power, i.e., strongly correlated with the top quark spin [39]. Consequently, in case of right handed hadronic decaying top quark, the $d$-type quark is expected to be more energetic compared to the $b$ and $u$ quarks, which are not as boosted and hence widely separated. On the contrary, the $b$ and $u$ quarks are more boosted, and thus less separated in case of left handed hadronic top quarks [14]. For leptonically decaying right (left) handed top quarks, the lepton ($b$ quark) is more boosted compared to the $b$ quark. Thus one can exploit these features of top quark decay products to construct various polarization observables [11–14]. These polarization driven kinematic characteristics also manifest themselves in the jet images, which can thereby be used to train the CNN.

Following the same methodology described before, preprocessed top jet images are produced from the SUSY sample where polarization of the top quark is set to either left or right handed. Figure 5, presents the jet images for left (upper left) and right (upper right) handed hadronic top quarks. The corresponding track and neutral hadron component images are also shown in the middle and lower panels of the same figure, respectively. The subtle differences between the images corresponding to the left and right handed top jets can be understood following the above discussion on the correlation between the decay product momenta and the top quark handedness. This suggests that for right handed hadronically decaying boosted top jets, the left (weaker) spot is more diffused and connected to the right (brighter) spot, compared to left handed top jets. This can be attributed to the fact that the bright right spot is often formed by the leading $d$ ($b$ or $u$) like subjet, while the comparatively less bright left spot is due to the $b + u$ ($u + d$ or $b + d$) system for right (left) polarized top quarks. Hence, for right handed top jets, the two spots are more connected as the $u$ and $d$ like jets are strongly color connected among themselves. Similar patterns are also observed for the track component images.

Figure 6 presents the preprocessed jet images for left (top left) and right (top right) polarized leptonic top jets. The middle and lower panels show the images corresponding to the track and neutral hadron components respectively. Clearly, significant differences in the jet images are observed for the two polarization states of the top quark. It is to be noted here that the charged lepton plays the same role as the $d$ quark in hadronic top decays, and hence the same arguments discussed above can be applied here in order to understand the patterns in the images. Accordingly, the brighter right spot is more often formed by the lepton ($b$ quark) for right (left) handed top jets. Thus the track component of this spot is harder for the right handed case, whereas the neutral hadron component is softer.

The CNN is trained (tested) using about 1M/1M (115K/115K) left/right handed top jet images from the SUSY sample. This training is also evaluated on the $W'$ sample to validate its robustness.

A robust angular variable, namely $\cos\theta^\star$, is constructed from the momenta of the subjets inside the top fatjet. This variable has been demonstrated to be a powerful discriminator for distinguishing between left and right handed top quarks [14]. In this methodology, the main challenge is to identify the $d$-like subjet. Currently, ML based techniques are available to tag $b$-jets inside the boosted top jet [40], which can be used as a handle to find the $d$-like subjet. However, implementing such a technique is beyond the scope of the present analysis. We follow the procedure described in [14], where the two subjets whose invariant mass is closest to the $W$ boson mass are labeled as non-$b$ subjets, and the third subjet is labeled as $b$-like. Then comparing the invariant masses between the $b$-like subjet and the other two subjets (say, $j_1$ and $j_2$), the $d$-like subjet is identified. For instance, if $m_{bj_1} < m_{bj_2}$, then $j_1$ is identified as the $d$-like subjet and used to construct $\cos\theta^\star$ as,

$$\cos\theta^\star = \frac{\vec{j}_t \cdot \vec{j}'_d}{|\vec{j}_t||\vec{j}'_d|}. \tag{4}$$

Here $\vec{j}_t$ is the momentum of reconstructed top jet in the lab frame, and $\vec{j}'_d$ is the momentum of the $d$-like subjet in the top rest frame. For the leptonic case, the lepton energy fraction

$$z_\ell = E_\ell/E_t, \tag{5}$$

acts as a sensitive polarimeter [10]. Here $E_\ell$ and $E_t$ are the energies of the lepton track and the top jet respectively, in the lab frame. We use the following procedure to identify the lepton track. The top jet constituents are exclusively clustered into two $k_T$ subjets. The mini-isolation [41] is computed for the leading (highest $p_T$) track inside each subjet. The track with smaller mini-isolation is identified as the lepton track. It is found that for about 80% cases this track matches with the MC generator level lepton.

The ROC curves corresponding to $\cos\theta^\star$ (dashed lines) and the CNN classifier (solid lines) are presented in Fig. 7 (left), for left and right handed hadronic top jets from the SUSY sample (in red). The performance of the CNN training when evaluated on top jets from the $W'$ sample is also shown (in blue). The analogous ROC curves for the leptonic top jets are presented in the right plot of the same figure. Evidently, the CNN classifiers outperform the kinematic variables ($\cos\theta^\star$, $z_\ell$), for both leptonic and hadronic top jets.

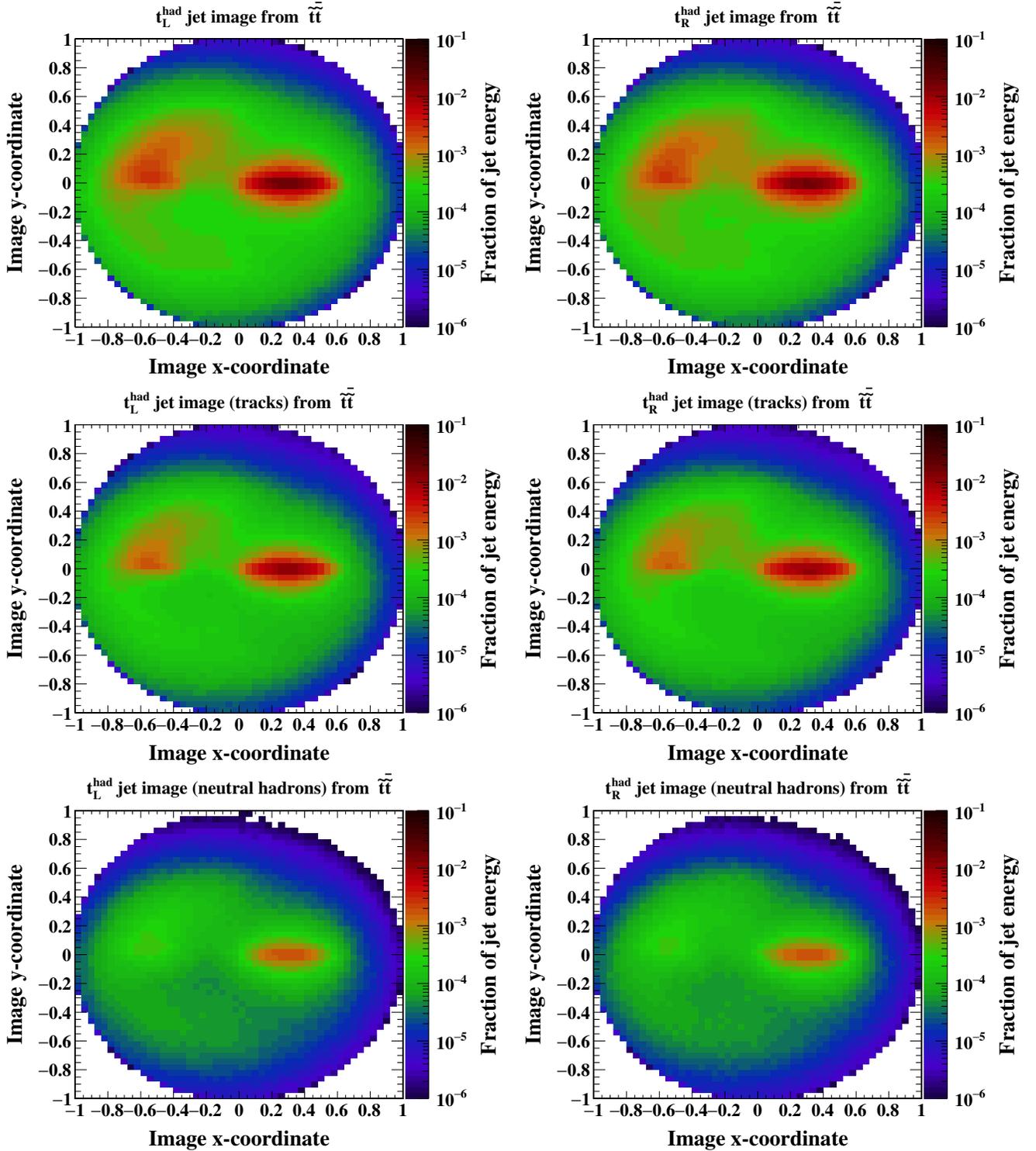

FIG. 5. The upper row presents the images of left-handed (left) and right-handed (right) hadronic top jets from $\tilde{t}_1$-pair events. The corresponding track (neutral hadron) component images are shown in the middle (lower) row.

The impact of the polarization sensitive variable is estimated using an experimentally measurable quantity, namely asymmetry, defined as follows.

$$A_v^P = \frac{N_{v>c} - N_{v<c}}{N_{v>c} + N_{v<c}}. \tag{6}$$

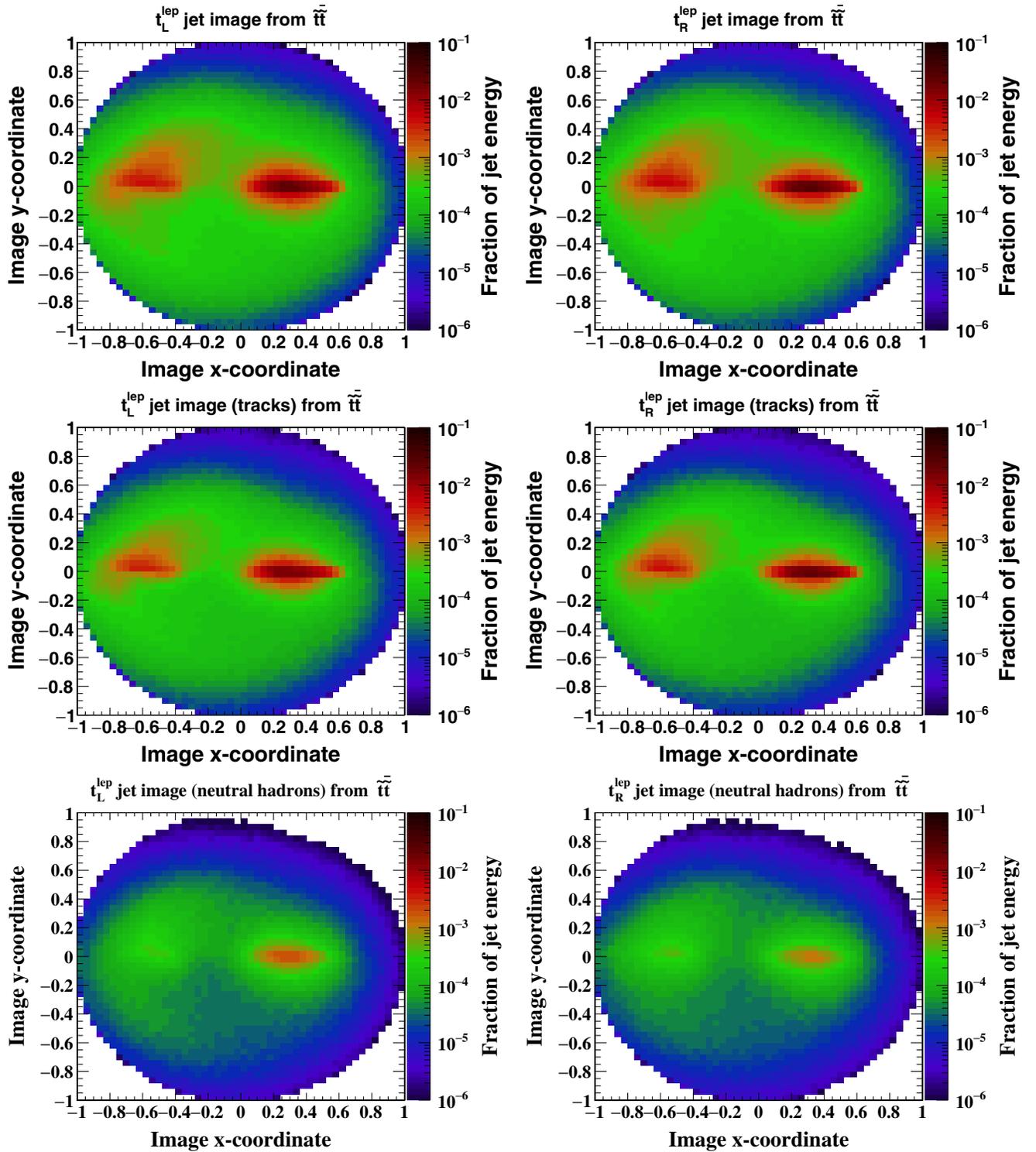

FIG. 6. The top row presents the images of left-handed (left) and right-handed (right) leptonic top jets from $\tilde{t}_1$-pair events. The corresponding track (neutral hadron) component images are shown in the middle (lower) row.

Here $N_{v>c}$ is the number of top jets subject to the condition that its polarization discriminator $v$ ($\equiv \cos\theta^\star$, $z_\ell$ or CNN classifier) is greater than a given threshold $c$, and $N_{v<c}$ is defined similarly. The superscript $P$ refers to the polarization composition of the top jets in a given sample. Recall that we consider only the two extreme compositions (entirely left or right handed) in this study. The magnitude of the difference $D_v = |A_v^L - A_v^R|$ is a measure of the maximal sensitivity of $v$ to the top quark polarization. Needless to say, this difference will be very small if $v$ is not very sensitive to polarization, and large otherwise. The optimum value of $c$ for a given discriminator $v$, is the point at which $D_v$ is maximum. Naturally, the most sensitive polarization discriminator is decided by comparing the peak values of $D_v$. The asymmetries and their differences are presented in Fig. 8 for hadronic (left) and leptonic (right) top jets from the SUSY sample. It is evident from the peak value of $D_v$, that the CNN classifier is $\approx 2$ times more sensitive compared to $\cos\theta^\star$ for hadronic top jets, and $\approx 1.3$ times more sensitive than $z_\ell$ for the leptonic case.

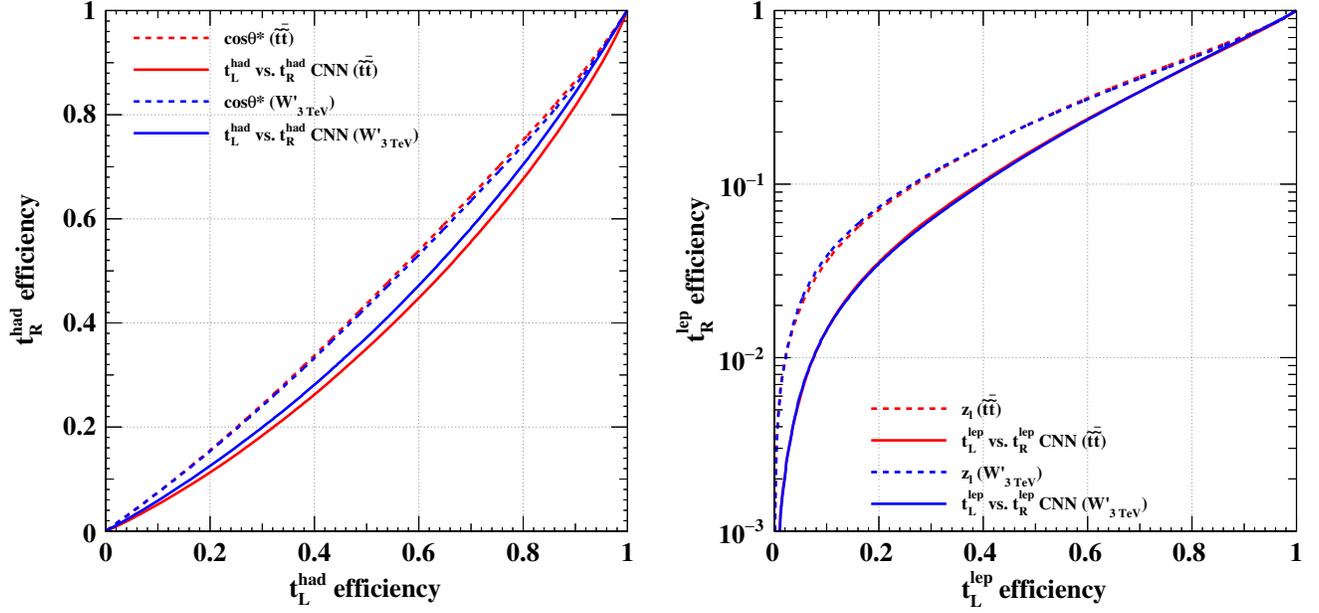

FIG. 7. The left plot shows the discrimination (ROC curve) between left-handed and right-handed hadronic top jets using $\cos\theta^*$ (dashed) and the CNN classifier (solid) in $\tilde{t}_1$-pair (red) as well as $W'$ (blue) events. Note that the CNN trained on top jets from $\tilde{t}_1$-pair events has been evaluated on top jets from $W'$ events. The right plot shows the same for leptonic top jets, except that the $z_\ell$ variable is used instead of $\cos\theta^*$.

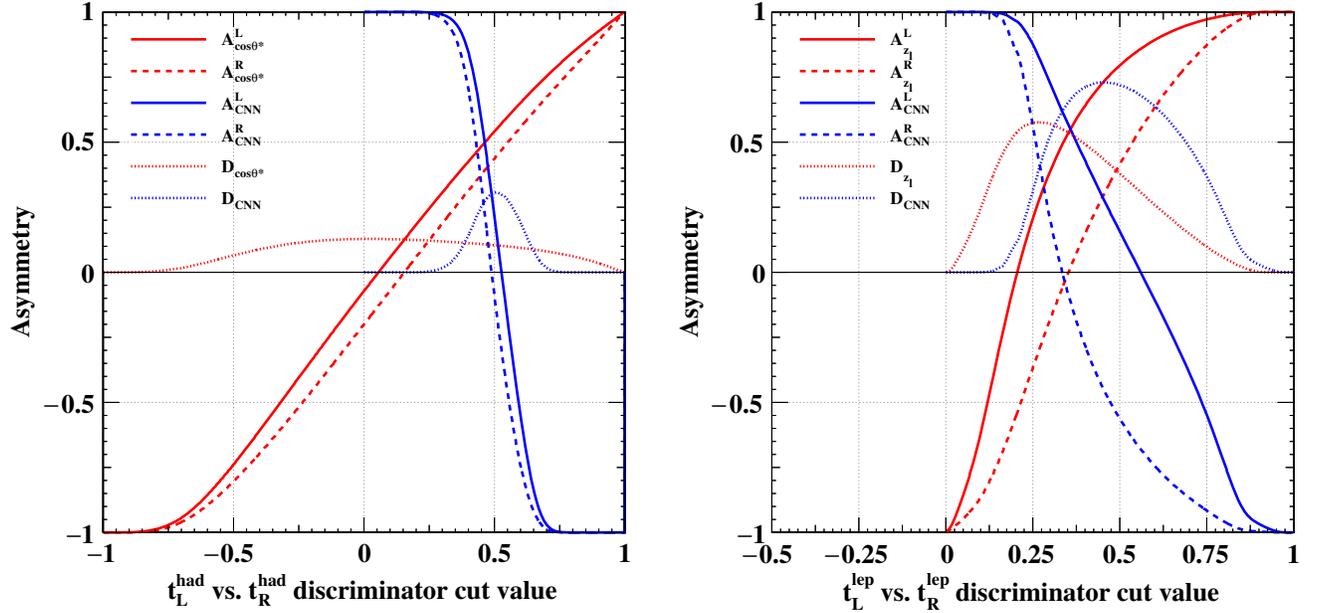

FIG. 8. The asymmetry variables [Eq. (6)] and their absolute differences between the left and right polarized cases are shown as a function of the discriminator threshold, corresponding to different polarization discriminators, using hadronic (left) and leptonic (right) top jets from the SUSY sample.

## V. SUMMARY

The results of boosted top tagging performances in hadronic and leptonic channels using jet images are presented, with an emphasis on the latter. The CNN is able to well identify leptonic top jets against a wide variety of backgrounds, namely, hadronic top, light flavor QCD, and leptonic and hadronic $W/Z$ jets. We have demonstrated that the tagging performance can be further improved by employing a BDT that uses the CNN classifier along with other high level inputs. The ROC curves show that top jet tagging efficiencies for the leptonic channel are far better than the hadronic mode, even when considering the backgrounds from $W/Z +$ jets events. We believe this to be a new and interesting observation, having promising implications in the context of LHC experiments. It is to be noted that no dedicated lepton identification is required when tagging leptonic top jets using images, which is a great advantage of this technique as nonisolated leptons inside a boosted jet can be very challenging to identify reliably.

The performance of the image based method in distinguishing between the two polarization states of the top quark has also been presented. This polarization measurement strategy is compared with two other kinematic polarimeter variables, namely $\cos\theta^\star$ and $z_\ell$ (for the hadronic and leptonic channels, respectively), using an asymmetry variable which is a measure of how sensitive a given polarimeter variable is to a change in the polarization composition. It is observed that the CNN classifier is more sensitive to polarization than the aforementioned kinematic polarimeters, and thus can be used to measure and distinguish between polarization states better and more reliably.

## ACKNOWLEDGMENTS

The authors are thankful to Suman Chatterjee (affiliated to Hochenergiephysik (HEPHY), Vienna) for useful discussions. It is to be noted that A. V. is an ex-member of TIFR.